# COMMUNICATION

# Efficient Growth and Characterization of One-Dimensional Transition Metal Tellurides Inside Carbon Nanotubes

Naoyuki Kanda,[a, b] Yusuke Nakanishi,*[a] Dan Liu,[c,d] Zheng Liu,[e] Tsukasa Inoue,[b] Yasumitsu Miyata,[a] David Tománek,[c] and Hisanori Shinohara[b]

**Atomically thin one-dimensional (1D) van der Waals wires of transition metal *mono*chalocogenides (TMMs) have been anticipated as promising building blocks for integrated nanoelectronics. While reliable production of TMM nanowires has eluded scientists over the past few decades, we finally demonstrated a bottom-up fabrication of MoTe nanowires inside carbon nanotubes (CNTs). Still, the current synthesis method is based on vacuum annealing of reactive $MoTe_2$, and limits access to a variety of TMMs. Here we report an expanded framework for high-yield synthesis of the 1D tellurides including WTe, an unprecedented family of TMMs. Experimental and theoretical analyses revealed that the choice of suitable metal oxides as a precursor provides useful yield for their characterization. These TMM nanowires exhibit a significant optical absorption in the visible-light region. More important, electronic properties of CNTs can be tuned by encapsulating different TMM nanowires.**

Nanocarbon materials – nanoscale substances comprised of carbon – have played a crucial role in modern materials science.[1-3] As the nanocarbon research matures, significant efforts have been directed towards creating 'post-nanocarbon' materials. Transition metal chalcogenides have been regarded as the promising candidates for the novel low-dimensional materials owing to their versatile chemistry and physics. Over the last decade, mono- or few layers of transition metal *di*chalcogenides have been widely recognized as the ideal platform to investigate 2D physics.[4] On the other hand, their 1D counterparts, which exist in a variety of morphologies including nanoribbons,[5] nanotubes,[6] and nanowires,[7,8] could exhibit the properties that can be differentiated from the 2D sheets as well as 1D nanocarbons. For instance, single units of transition metal *mono*chalcogenides (TMMs) – in which atomically thin wires are attracted with each other via van der Waals interaction – are 1D metallic wires,[8-10] and thus can serve as ultrathin channels for 2D integrated circuit. We wish to emphasize that unlike in an infinite 2D sheet, electronic states in NWs are quantized due to their finite width. This fact has been explored extensively in finite-width graphene nanoribbons (GNRs) with a fundamentally different conductance from an infinite graphene monolayer.[11] Similar effects are expected in the systems we study. Furthermore, isolated wires are believed to exhibit torsional motion not seen in the bulk, and thereby change their bandgap depending on their torsional angle,[12] allowing potential applications in electro-mechanical switching devices. Despite the growing interest, only a few kinds of isolated TMM nanowires (NWs) are currently available.[13,14] In particular, the isolation of environmentally unstable tellurides with chemical precision remains a significant challenge. In theory, isolated MoTeNWs could exhibit semiconducting properties that opposite the metallic sulfides and selenides, although their salient properties have never been fully verified by experiments.[15,16]

One promising way to produce the 1D tellurides is by using a template reaction via carbon nanotubes (CNTs), in which the assembly of the NWs proceeds in one direction.[17] Thermally and chemically robust CNTs have increasingly been used to produce unstable 1D materials such as atomic wires,[18,19] carbyne,[20,21] diamond nanowires,[22,23] GNRs,[24,25] and molecular arrays.[26,27] Besides acting as a template, CNTs also provide a permanent protection for the exposed edges of NWs from the ambient. Perfect CNTs are chemically as inert as defect-free graphene and cannot be destroyed or removed easily. On the other hand, similar to graphene, carbon nanotubes are rather unreactive. Indeed, we have recently demonstrated the

[a.] *Department of Physics, Tokyo Metropolitan University, Tokyo 192-0397, Japan. E-mail: naka24ysk@gmail.com*
[b.] *Department of Chemistry, Nagoya University, Nagoya 464-8602, Japan.*
[c.] *Department of Physics and Astronomy, Michigan State University, East Lansing, Michigan 48824, United States.*
[d.] *Theoretical Division, Physics and Chemistry of Materials, Los Alamos National Laboratory, Los Alamos, New Mexico 87545, United States.*
[e.] *National Institute of Advanced Industrial Science and Technology, Nagoya 463-8560, Japan.*

Electronic Supplementary Information (ESI) available: Materials, experimental and theoretical techniques, synthesis conditions, chemical and spectroscopic analysis. See DOI: 10.1039/x0xx00000x

successful growth of MoTeNWs inside CNTs by a vacuum annealing of bulk MoTe$_2$.[28] We found that CNTs do not form covalent bonds to the enclosed NWs and thus should not modify their properties. Atomic-level transmission electron microscopy revealed that MoTeNWs show discontinuous torsion, significantly distinct from continuous twisting of MoSNWs.[29] These unique mechanical properties suggest that MoTeNWs and the related 1D tellurides are distinguishable from other TMMs. In order to further explore their properties, a method for their high-yield growth is needed to be developed. Here we demonstrate efficient production of MoTeNWs inside CNTs and its expansion to WTeNWs, an unprecedented member of the TMM family. Through experimental and theoretical studies, we found that the production yields are highly enhanced by choosing suitable metal oxides as a precursor. Furthermore, their high-yield growth enabled us to study their electronic and optical properties. Our CNT-templated growth of MoTe- and WTeNWs will usher in further experiments to shed light on their unexplored properties and potential applications.

A schematic image of the present strategy for the growth of MoTeNWs is shown in Fig. 1a. We chose MoO$_2$ as precursor in this work. We have previously noted the possibility that molybdenum oxides act as intermediates in the reaction that uses MoTe$_2$ as a precursor. Early literature demonstrated that vacuum annealing of MoTe$_2$ results in the formation of non-volatile Mo solid with Te vapors,[30] suggesting that other chemical species serve as the direct source of Mo instead of bulk MoTe$_2$. High-resolution X-ray photoelectron spectroscopy (XPS) revealed that precursor MoTe$_2$ contains several molybdenum oxides including highly volatile MoO$_2$.[28] Thus, we hypothesized that MoO$_2$ is the direct precursor to MoTeNWs.

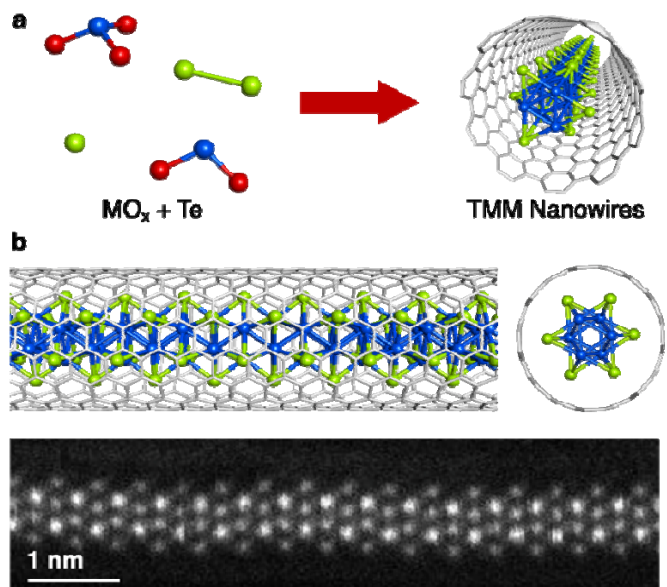

**Fig. 1** (a) Schematic of the CNT-templated reaction for the formation of MoTeNWs. (b) A structural model and atomic-resolution HAADF-STEM image of an individual MoTeNW@CNT.

Fig. 1b displays a representative high-angle annular dark-field scanning transmission electron microscopy (HAADF-STEM) image of the as-produced samples. The atomic structure of the present NWs is similar to the products from MoTe$_2$, which have a quasi-1D structure consisting with Mo$_3$Te$_3$ triangles stacked alternately along the $c$-axis of CNTs. The positions of XPS peaks are consistent with those of MoTeNWs inside CNTs (Fig. S1a). Chemical analysis by energy-dispersive X-ray spectroscopy (EDS) also suggests that the NWs are composed of Mo and Te atoms (Fig. S1b). All the results indicate direct growth of MoTeNWs from a mixture of MoO$_2$ and Te.

Importantly, this reaction utilizing MoO$_2$ as a precursor significantly enhanced the yield of MoTeNWs. Fig. 2a shows Raman spectra of the samples obtained in the optimized conditions from the two different precursors (Fig. S2). Raman spectrum of MoTeNWs exhibits a characteristic peak located at 251 cm$^{-1}$, which is assigned to A$_g$ radial breathing mode (RBM) of Mo cores (Fig. 2b).[28] As for the samples from MoO$_2$ with Te, the intensity ratio of the A$_g$ peak to G-band of CNTs is 19.4%, which is approximately 20 times larger than the corresponding value (~1%) from bulk MoTe$_2$ containing the equivalent amount of Mo atoms (Fig. S3). The enhanced A$_g$ peak of MoTeNWs revealed an improvement for their production yields. The high-yield synthesis of MoTeNWs was further confirmed by HAADF-STEM images (Fig. 2c, d).

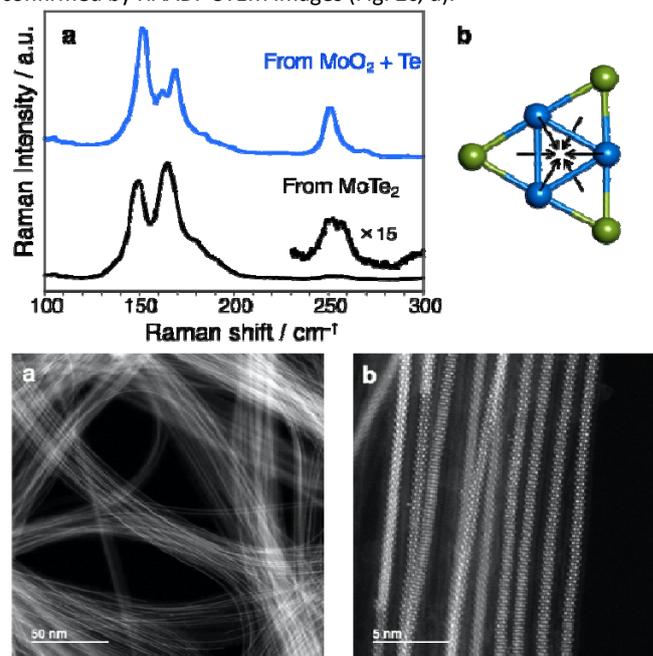

**Fig. 2** (a) Raman spectra of MoTeNW@CNTs *via* vacuum annealing of two different precursors (MoTe$_2$ vs MoO$_2$ + Te) at the 633 nm excitation. (b) Schematic of the atomic vibrations for the Raman modes at 251 cm$^{-1}$. (c, d) HAADF-STEM images of MoTeNW@CNTs.

In parallel to the experiments, we performed density functional theory (DFT) calculations to investigate the growth mechanism. The present reaction was rationalized by calculating Δ$E$ for the following reaction:

$x$MoO$_2$+$y$MoO$_3$+($x$ + 1.25$y$)Te$_2$
$\rightarrow$ (Mo$_3$Te$_3$)$_3$+(2$x$ + 3$y$)/2TeO$_2$–Δ$E$ ($x$ + $y$ = 9)     (1)

where Mo$_3$Te$_3$ represents half the unit cell in an individual MoTeNW. The formation of TeO$_2$ as a side product was verified by XPS spectra of the residues attached on the surfaces of CNTs.[28] The enthalpy changes for various combinations of (*x*, *y*) are summarized in Table 1. Our DFT computation reveals that the reaction in $x \geqq 4$ is an exothermic process, and proceeds favorably in terms of thermodynamic energy. In particular, the enthalpy changes for the reaction starting with pure $MoO_2$ ($x$ = 9, $y$ = 0) is the largest (-16.8 eV). This reactivity can be explained by considering that the dissociation of three Mo–O bonds in $MoO_3$ requires higher energy.[31] These DFT results reveal that $MoO_2$ plays a key role in the reaction.

Table 1 Reaction energy in equation (1).

| x | y | ΔE (eV) |
|---|---|---|
| 4 | 5 | −0.245 |
| 5 | 4 | −3.55 |
| 6 | 3 | −6.86 |
| 7 | 2 | −10.1 |
| 8 | 1 | −13.5 |
| 9 | 0 | −16.8 |

Our calculations also indicate that CNTs play a crucial role in the NWs formation. Mo and Te atoms are stabilized on monolayer graphene due to charge transfer between the atoms and the sheet (Fig. S4), suggesting that the molecules easily get adsorbed on the surfaces of CNTs. Moreover, the electron transfer from $MoO_2$ to the graphene causes a substantial weakening of the Mo–O bonds (Fig. S5), and thereby the absorbed $MoO_2$ can readily react with Te species.

We further calculated ΔE for the sequential extension of $Mo_3Te_3$ triangles:

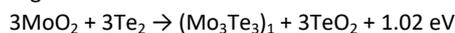
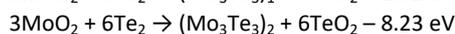
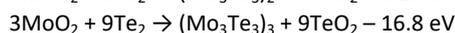

$3MoO_2 + 3Te_2 \rightarrow (Mo_3Te_3)_1 + 3TeO_2 + 1.02$ eV
$3MoO_2 + 6Te_2 \rightarrow (Mo_3Te_3)_2 + 6TeO_2 - 8.23$ eV
$3MoO_2 + 9Te_2 \rightarrow (Mo_3Te_3)_3 + 9TeO_2 - 16.8$ eV

A series of reaction enthalpy calculations suggests that the assembled triangles are energetically preferred to isolated ones. This thermodynamic stability favors spontaneous assembly of the NWs. Indeed, the overall length of MoTeNWs reaches a submicron scale, which far exceeds that of typically reported 1D materials inside CNTs.

A possible mechanism for the NWs formation is proposed in Fig. 3. Sublimed $MoO_2$ and $Te_2$ molecules are first adsorbed on the outer walls of CNTs. The absorbed molecules can diffuse freely along the surfaces and finally enter the channels due to their stabilization in the inner void.[32,33] As a consequence, the molecules readily react with each other to assemble within CNTs. The self-assembly process occurs repeatedly, resulting in the formation of elongated MoTe NWs. Quasi-1D CNTs serve as a template that prevents the formation of branched structures and bundles by restricting the direction of the chemical reaction.

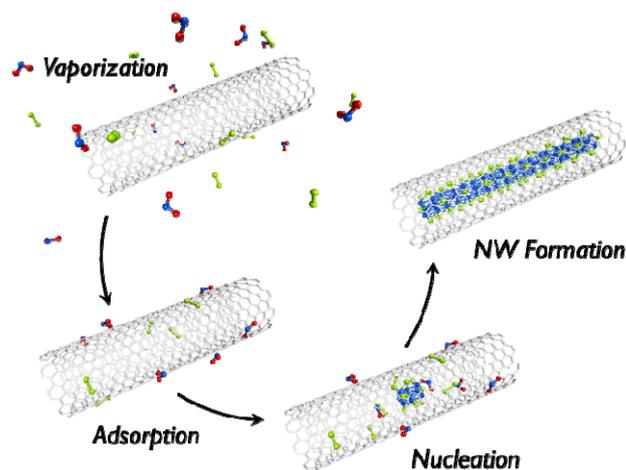

**Fig. 3** Possible pathway for growth of MoTeNWs inside CNTs.

Based on the present methodology, we further successfully created WTeNWs that have never previously been found in the bulk. A typical HAADF-STEM image of the NWs produced from a mixture of $WO_3$ and Te is displayed in Fig. 4a. Chemical composition of the NWs was characterized by the EDS in Fig. 4b, showing that the resultant NWs are composed of W and Te atoms. The axial lattice constant and diameter are measured to be 0.45 and 0.48 nm (Fig. 4c, S6), which are equivalent to the corresponding values of MoTeNWs and SnSeNWs.[34] The filling yield of WTeNWs is comparable to that of MoTeNWs. In HAADF-STEM images, heavier W atoms ($Z$ = 74) look brighter than Mo atoms ($Z$ = 42) due to the strong electron scattering (Fig. S7). We found that isolated WTeNWs discontinuously twist similar to MoTeNWs (Fig. S8), which is significantly distinct from MoS,[29] $NbSe_3$,[35] and $HfTe_3$NWs.[36] Chemical states of WTeNWs confined within CNTs are characterized by XPS. Fig. 4d shows the W 4*f* and Te 3*d* core-level XPS spectra, where the peaks at 31.1 and 572.4 eV are assigned to W $4f_{7/2}$ and Te $3d_{5/2}$ of WTeNWs. Compared to $WTe_2$ layers (31.6 and 572.7 eV), the core-level binding energies for W $4f_{7/2}$ and Te $3d_{5/2}$ show downward shifts by 0.5 and 0.3 eV, respectively. The blue-shifted XPS core-levels are in accord with the peak shift for MoTeNWs.

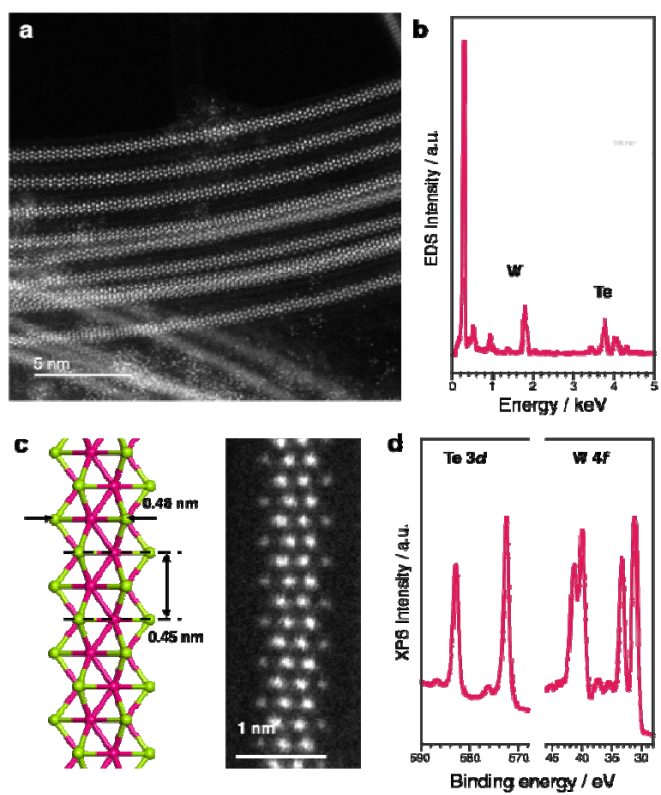

**Fig. 4** (a) A typical HAADF-STEM image and (b) EDS analysis of WTeNW@CNTs. (c) A typical HAADF-STEM image and atomic structure of an individual WTeNW. (d) XPS spectra of W 4f and Te 3d core-levels of WTeNWs confined in CNTs.

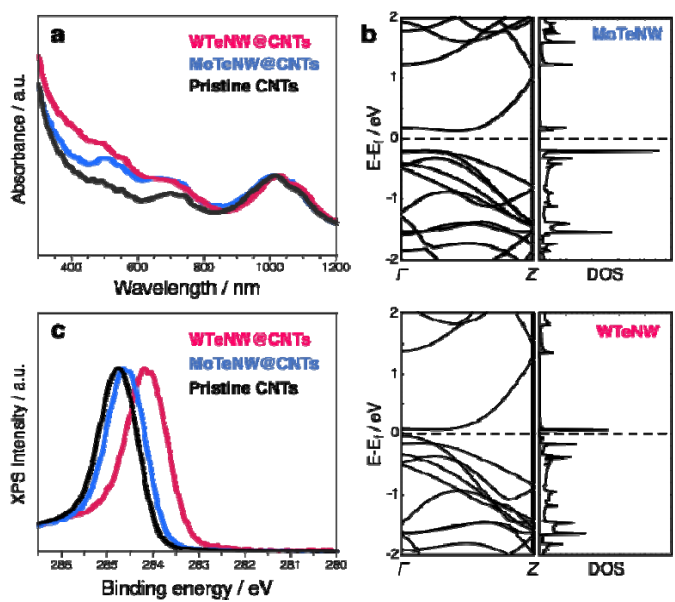

**Fig. 5** (a) Optical absorption spectra of MoTe- (blue), WTeNW@CNTs (pink), and pristine CNTs (black). (b) Bandstructures and DOS of isolated MoTe- and WTeNWs. (c) XPS spectra of C 1$s$ core-levels of MoTe- (blue), WTeNW@CNTs (pink), and pristine CNTs (black).

Efficient growth of these NWs has enabled us to compare their optical and electronic properties. Optical absorption spectra of the products are presented in Fig. 5a. Although the absorption spectra are dominated by the features of template CNTs rather than inner NWs, their spectra commonly show a significant band in the range of 450–600 nm. In order to obtain information on the absorption bands, the electronic structures and density of states (DOS) of isolated MoTe- and WTeNWs were computed by means of DFT calculation. As displayed in Fig. 5b, both of the isolated NWs can be narrow-gap semiconductors in contrast to metallic MoS- and MoSeNWs.[13] Even though DFT is known to underestimate bandgaps, it still can provide useful information about the electronic structure near the Fermi level. Compared to MoTe, the bottom of the conduction band of WTeNWs at the Γ point is lowered, and the band is nearly flat. DFT-based bandgaps between the highest occupied and the lowest unoccupied states of MoTe- and WTeNWs at the Γ point are estimated to be 0.39 and 0.24 eV, respectively. As expected, these values lie well below the absorption bands (2.1~2.8 eV). On the other hand, the energy gaps between the highest occupied and the second lowest unoccupied states at Γ are estimated to be 1.44 eV for MoTeNWs and 1.51 eV for WTeNWs. The DOS diagrams in Fig. 5b suggest a possibility of corresponding inter-band transitions between van Hove singularities. There is little difference in the transition energies between MoTe- and WTeNWs, which also agrees with the experimental results.

Fig. S9 compares Raman spectra of MoTe- and WTeNW@CNTs. As illustrated in Fig. 2a, b, MoTeNWs confined in CNTs exhibit a prominent peak at 251 cm$^{-1}$ originating from RBM mode of Mo cores. In contrast, there is a noticeable peak at 190 cm$^{-1}$ in the spectrum of WTeNW@CNTs. Given that the frequency is inversely proportional to the square root of mass ($f \propto m^{-1/2}$), the peak might be attributed to RBM of W cores. However, the peak may be assigned the shifted RBMs of CNTs due to the interaction with WTeNWs. Additional experiments are needed to figure out the origin of the Raman mode.

Further XPS analyses revealed that the charge transfer occurs between CNTs and the inner NWs. Fig. 5c shows C 1$s$ core-level XPS spectra of MoTeNW@CNTs, WTeNW@CNTs, and pristine CNTs. The C 1$s$ core-level is shifted downward by 0.6 eV after the encapsulation of WTeNWs, which is larger in comparison to MoTeNWs (~0.3 eV). The blue shift of C 1$s$ core-level is considered as a result of electron transfer from CNTs to the NWs.[37,38] Given that the filling yields of these NWs are comparable, WTeNWs exhibit the stronger affinity towards CNTs. The difference in the charge transfer can be understood by considering the difference in electronegativity between Mo and W atoms. The Pauling electronegativity of Mo ($\chi$ = 2.16) is almost equivalent to that of Te ($\chi$ = 2.1).[39] When Mo atoms are replaced by more electronegative W ($\chi$ = 2.36), the electron density in W–Te bonds is further pulled to W atoms. As a consequence, the positively charged Te atoms interact more strongly with inner walls of CNTs, thereby inducing stronger $p$-type doping of CNTs. This doping should shift the Fermi level and make all filled nanotubes conducting. Indeed, the electrical and thermal transport properties of CNTs are drastically modulated by encapsulated molecules,[40,41] leading to their application in electrical and thermoelectric devices.

We expect that confinement of the heavy-metal NWs will induce a significant modulation of transport properties of individual CNTs. A conclusive confirmation of this claim poses, however, a significant challenge. The main reason is that quantum conductance measurements are extremely demanding due to the difficulty to precisely position a nanotube and provide low-resistance ohmic contacts. These issues are currently unsolved and topics of high-priority research.

## Conclusions

In conclusion, we reported an expanded framework for high-yield synthesis of MoTe- and WTeNWs with chemical precision by utilizing the CNT-templated reaction. Vacuum annealing of suitable metal oxides with Te results in efficient growth of MoTe- and WTeNWs inside CNTs. These 1D products exhibit intense optical absorption peaks in the visible-light range. Also, the inner NWs interact strongly with the outer carbon sheaths, which thereby induce *p*-type doping of CNTs. Direct measurements of their transports are now underway in the present laboratory.

## Authors contributions

Y.N. conceived the idea and directed the project. N.K. conducted the sample preparation, spectral characterization, and preliminary TEM observation. Z.L. performed the STEM experiments. D.L. and D.T. performed DFT calculations related to growth and characterization. T.I. and H.S. conducted XPS experiments and analysis. N.K., Y.N., and Y.M. carried out absorption measurements and analyses. Y.N. prepared the manuscript with feedback from the other authors. All authors have given approval to the final version of the manuscript.

## Conflicts of interest

There are no conflicts to declare.

## Acknowledgements

We appreciate Shivani Shukla (Carnegie Mellon University) for a fruitful discussion. We also thank Dr. Yohei Yomogida (Tokyo Metropolitan University) for technical assistance. This work was financially supported by KAKENHI (18K14088 and 20H02572 to Y.N.). Y.N. also acknowledges Nagoya University–AIST alliance project 2019, Murata Science Foundation 2019, and JKA and its promotion funds from KEIRIN RACE. D.L. and D.T. acknowledge financial support by the NSF/AFOSR EFRI 2-DARE grant number EFMA-1433459. Computational resources have been provided by the Michigan State University High Performance Computing Center

# Supplementary Information

## Efficient Growth and Characterization of One-dimensional Transition Metal Tellurides Inside Carbon Nanotubes

Naoyuki Kanda,[a,b] Yusuke Nakanishi,[*a] Dan Liu,[c,d] Zheng Liu,[e] Tsukasa Inoue,[b] Yasumitsu Miyata,[a] David Tománek,[c] and Hisanori Shinohara[b]

[a.] Department of Physics, Tokyo Metropolitan University, Tokyo 192-0397, Japan.
[b.] Department of Chemistry, Nagoya University, Nagoya 464-8602, Japan.
[c.] Department of Physics and Astronomy, Michigan State University, East Lansing, Michigan 48824, United States.
[d.] Theoretical Division, Physics and Chemistry of Materials, Los Alamos National Laboratory, Los Alamos, New Mexico 87545, United States.
[e.] National Institute of Advanced Industrial Science and Technology (AIST), Nagoya 463-8560, Japan.

**Address correspondence to:**
naka24ysk@gmail.com (Y.N.)

## Materials & Methods

**Preparation of MoTeNWs.** High-quality arc-discharge CNTs (Meijo Nano Carbon Co. Ltd.) were employed as templates for the NWs formation. Closed ends of CNTs were opened by an oxidation treatment, during which the temperature is increased to 500 °C in 6h in air. The as-prepared CNTs were degassed for 1h under a vacuum of $10^{-7}$ Torr. Typically, 0.3 mg of the open-ended CNTs was sealed in a straight quartz tube under vacuum ($10^{-7}$ Torr) with 3 mg of $MoO_2$ and 3 mg of Te, and then heated at 1100 °C for 48h.

**Preparation of WTeNWs.** Typically, 0.3 mg of the open-ended CNTs was sealed in a straight quartz tube under vacuum ($10^{-7}$ Torr) with 1.5 mg of $WO_3$ and 3 mg of Te, and then heated at 900 °C for 48h.

**Raman Spectroscopy.** Raman scattering measurement was carried out by means of the invia confocal Raman systems equipped with a microscope (HORIBA JOBIN Yvon). The laser spot size was ɸ4 μm and the laser power was 400 μW. The excitation wavelength was 633 nm (He-Ne Laser).

**$C_s$-corrected STEM observation.** $C_s$-corrected HAADF-STEM images were taken by using a JEM-ARM200F ACCELARM (cold FEG) equipped with a CEOS $C_s$ corrector (ASCOR system), operated at 120 keV. The scan rate was 38 microseconds (38 μs) per pixel.

**XPS Spectroscopy.** XPS analyses were performed by using an ESCALAB 250Xi (Thermo Fisher Scientific Inc.) by using monochromatic Al X-ray source. Instrument-base pressure was $7 \times 10^{-10}$ Torr. High-resolution XPS spectra were collected using an analysis 650 μm in diameter and 50 eV pass energy. Charge neutralization system was employed for all analyses by flood gun with Ar gas.

**Absorption Spectroscopy.** Optical absorption spectra were measured on Shimadzu UV-3600 spetrophotometer with a path of 10 mm.

**DFT Simulations.** We have investigated the equilibrium geometry and structural stability of a series of molecules including molybdenum oxides, tellurium oxide $TeO_2$, and $Mo_xTe_x$ (*x*=3,6,9). The interaction between $MoO_2$ and graphene was calculated using *ab initio* DFT as implemented in the VASP code.[1-3] We studied these molecules in a periodic array separated by a vacuum region in excess of 25 Å. The 2D system of $MoO_2$ on graphene was represented by a periodic array of layers separated by a vacuum region of >20 Å. We employed projector-augmented wave (PAW) pseudopotentials[4,5] and the Perdew–Burke–Ernzerhof (PBE) exchange-correlation functional.[6] The Brillouin zone of the conventional unit cell of 0D and 2D structures has been sampled by a uniform k-point grid.[7] The specific sampling was 1×1×1 for isolated molecule optimization, and 3×3×1 for $MoO_2$ on graphene. We employed 600 eV as the electronic kinetic energy cutoff for the plane-wave basis and a total energy difference between subsequent self-consistency iterations below $10^{-5}$ eV/atom as the criterion for reaching self-consistency. All the geometries have been optimized using the conjugate-gradient method,[8] until none of the residual Hellmann–Feynman forces exceeded $10^{-2}$ eV/Å.

## 2. Optimization of the reaction temperature

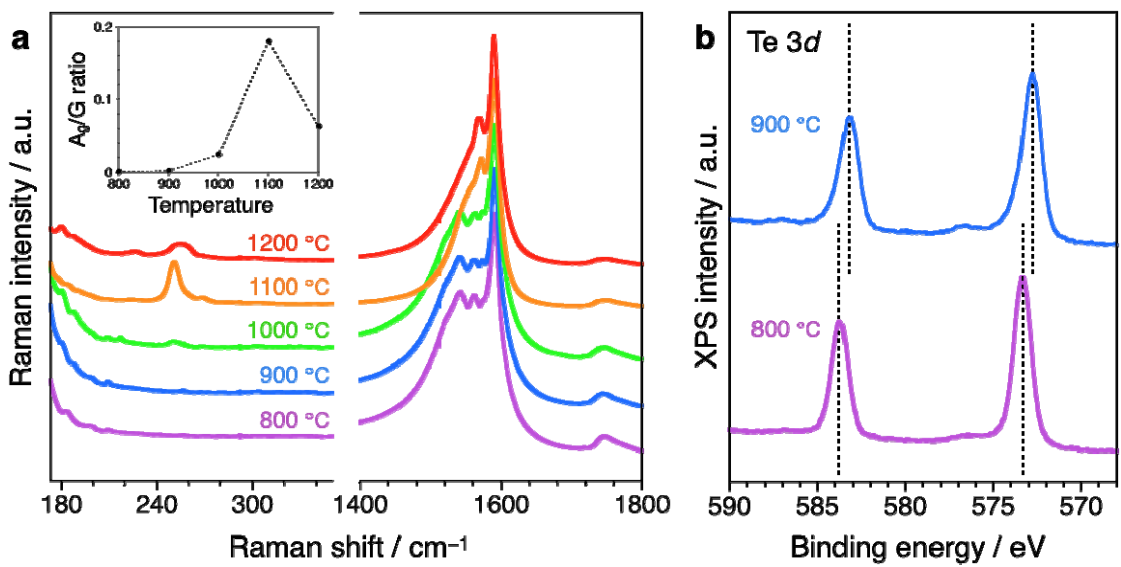

**Fig. S1** (a) Raman spectra of the products annealed at different temperatures (800–1200 °C), excited at the laser wavelengths of 633 nm. Inset shows the relative Raman intensity at 251 cm$^{-1}$, which is defined as the ratio of Raman intensity at 251 cm$^{-1}$ (due to A$_g$ mode of MoTeNWs) to G-band of CNTs at 1583 cm$^{-1}$, plotted as a function of annealing temperature. (b) XPS Te-3$d$ core-levels spectra of the products obtained at 800 and 900 °C.

The formation of MoTeNWs is strongly dependent on the annealing temperature. Raman spectroscopy was performed on the products, which were annealed at a series of different temperatures (Fig. S1a). The Raman signal at 251 cm$^{-1}$ associated with A$_g$ mode of MoTeNWs arises as the starting materials are annealed in vacuum above 900 °C. As shown in the inset of Fig. S1a, the intensity of this Raman feature (relative intensity of this peak to G-band of CNTs) increases with increasing temperature and reaches its maximum at 1100 °C, and decreases with the decomposition of the NWs above the temperatures (~1200 °C). The temperature dependence of the Raman signal suggests that annealing of the reactants at 1100 °C produces MoTeNWs in high yield. Further XPS analysis supported that the evolution of chemical reactions occurs between 800 and 900 °C. As displayed in Figure S1b, XPS spectrum of the products obtained at 900 °C show the Te-3$d_{5/2}$ and Te-3$d_{3/2}$ peaks at binding energies of 572.6 and 583.0 eV, which are attributable to the formation of MoTeNWs inside CNTs. Compared to 900 °C, a 0.8 eV blueshift was observed at 800 °C. The shifted peak can be assigned to pure Te, suggesting that TeNWs are predominantly formed inside CNTs.

## 3. Chemical analysis by means of XPS and EDS

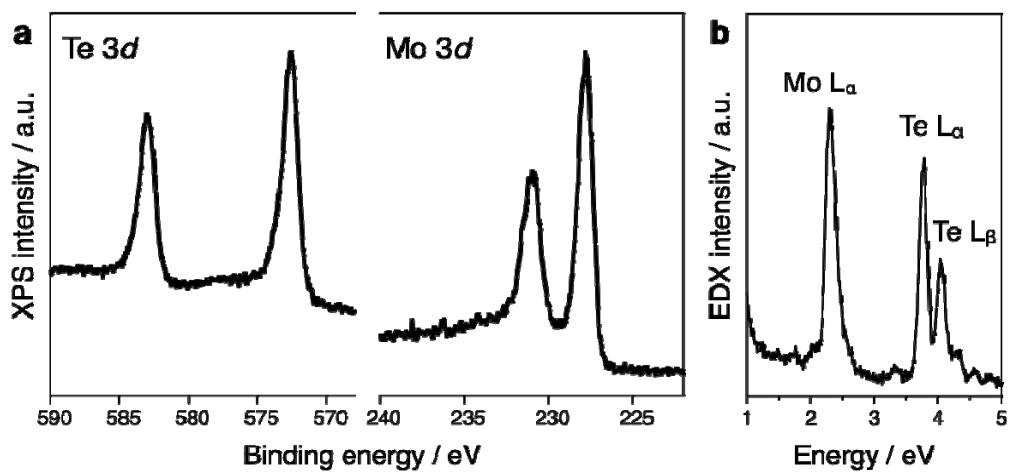

**Fig. S2** (a) XPS and (b) EDS spectra of MoTeNWs inside CNTs. As displayed in Figure S2a, Mo-3$d$ and XPS Te-3$d$ core-levels spectra of the present products show the Mo-3$d_{5/2}$, Mo-3$d_{3/2}$, Te-3$d_{5/2}$, and Te-3$d_{3/2}$ peaks at binding energies of 227.8, 231.0, 572.6 and 583.0 eV, respectively. The binding energies are in accordance with the values of the previously reported MoTeNWs encapsulated inside CNTs.[9] Quantitative chemical analyses by means of EDS yields 1:1 atomic ratio of Mo and Te.

## 4. Quantitative Raman Spectroscopy

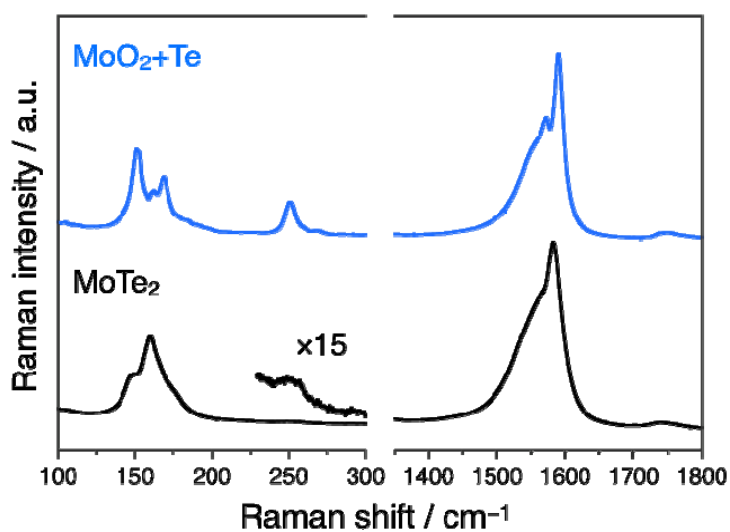

**Fig. S3** Expanded Raman spectra of the products obtained *via* vacuum annealing of the mixture of MoO$_2$ and Te (blue) and bulk MoTe$_2$ crystals (black). Notice that these two starting materials contain the same amount of Mo atoms (0.023 mmol). The sample preparations were carried out under the same conditions ($10^{-7}$ Torr, 1100 °C for 48h).

## 5. Additional Results of DFT Calculations

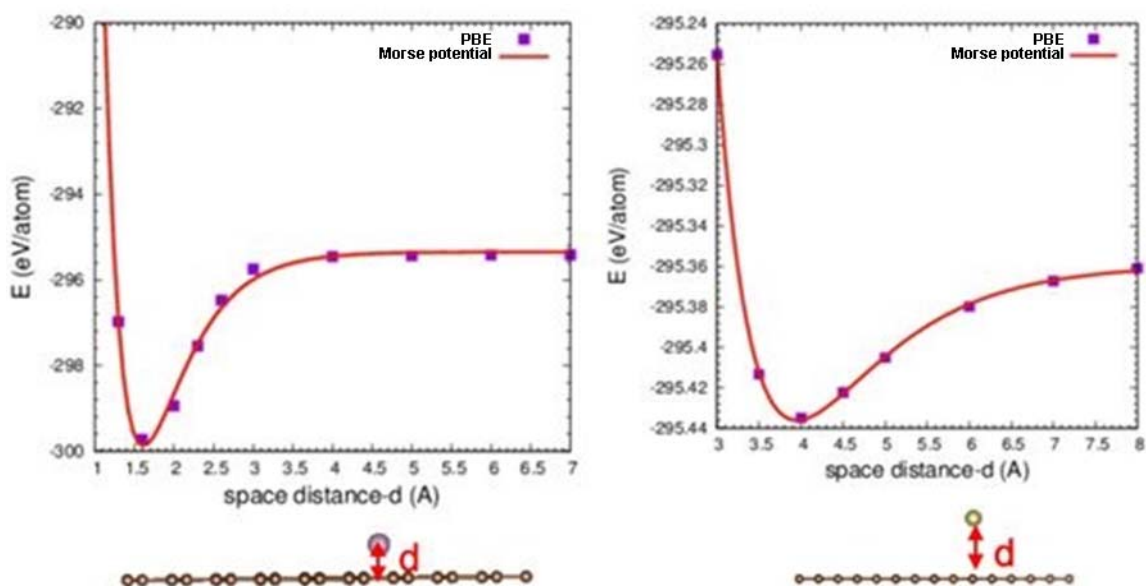

**Fig. S4** Stabilization energy and equilibrium geometry of (left) Mo and (right) Te atoms adsorbed on graphene as a function of the separation *d*.

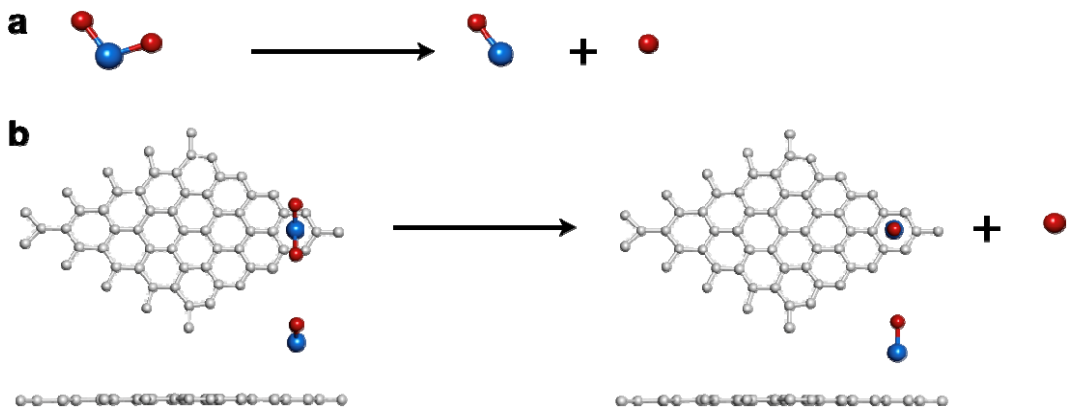

**Fig. S5** Breaking the Mo–O bond of the MoO$_2$ molecule (a) in free space and (b) on graphene.

In order to figure out the effect of CNTs on the chemical reactions, we have calculated the dissociation energy for a bond Mo–O in MoO$_2$ molecule without and with a graphene sheet (Fig. S5). All the geometries were fully optimized within appropriate symmetry constraints. The energy which is needed to break the Mo–O bond through the reaction shown in Figure S5a is estimated to be 9.684 eV. On the other hand, the Mo–O dissociation energy for MoO$_2$ molecule attached on the graphene is lowered to 7.747 eV. In the optimized structures in Fig. S5b, the Mo atom faces graphene and transfers 0.236 electrons to the sheet. As a consequence, the energy needed to break Mo–O bond is significantly reduced.

## 6. Further STEM information

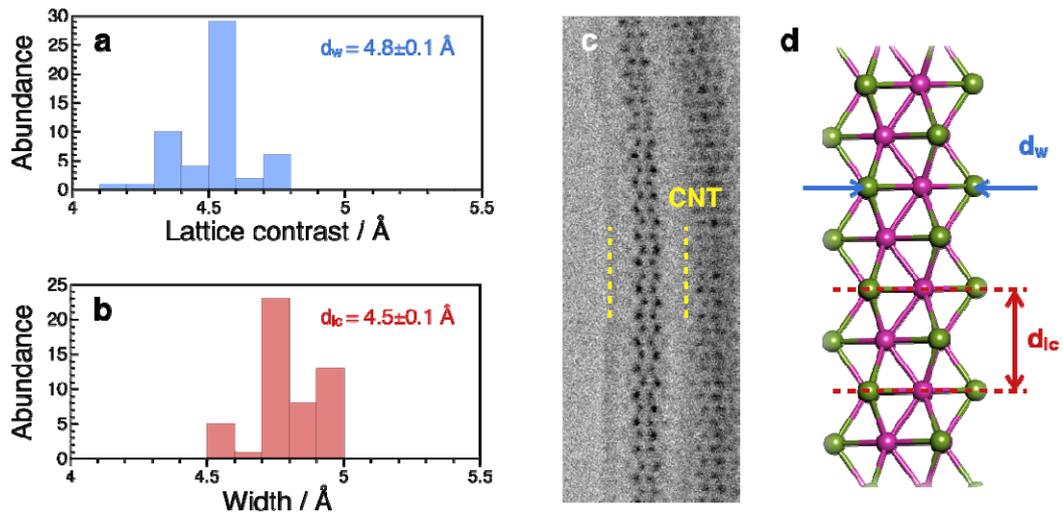

**Fig. S6** Distribution of (a) widths and (b) lattice parameters in WTeNWs. (c) A representative $C_s$-corrected TEM image and (d) atomic structure of an individual WTeNW confined within CNTs.

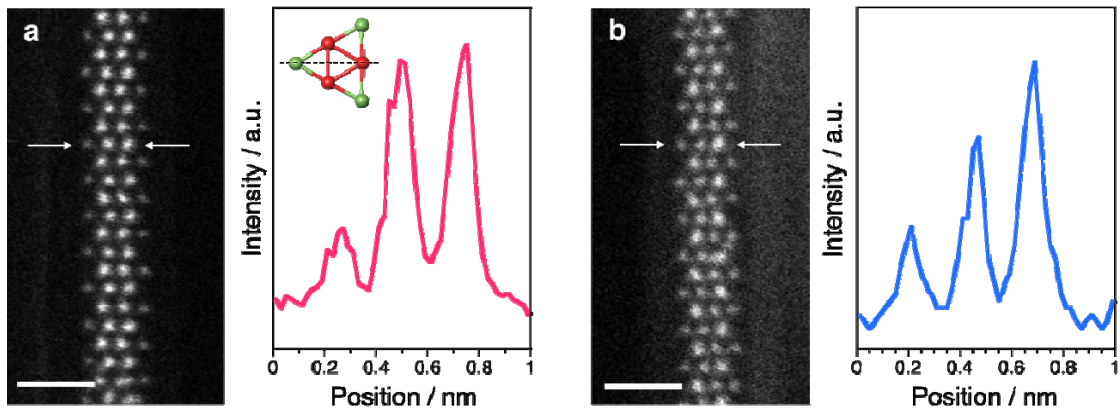

**Fig. S7** Typical HAADF-STEM images of (a) WTe- and (b) MoTeNWs confined within CNTs and line intensity profiles along the *X-X'* and *Y-Y'* directions in the corresponding images, respectively. Scale bar, 1 nm.

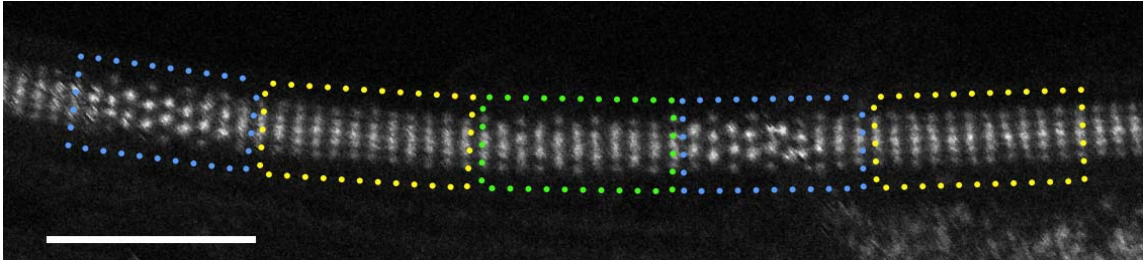

**Fig. S8** Torsional motions of an individual WTeNW confined within CNTs. Scale bar, 3 nm.

## 7. Raman signals of MoTe- and WTeNWs confined in CNTs

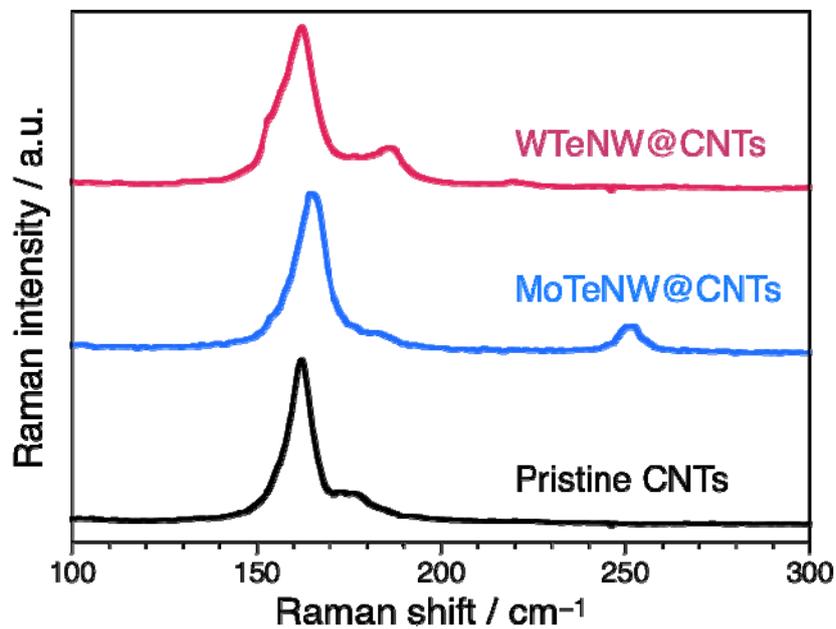

**Fig. S9** Raman spectra of WTeNWs@CNT (pink), MoTeNW@CNTs (blue) and pristine CNTs (black).